# Analysis of Queue Length Prediction from Probe Vehicles Problem with Bunch Arrival Headways

Gurcan Comert

*Assistant Professor*
*Dept. of Math and Computer Science*
*Benedict College*
*comertg@benedict.edu*

Stacey Franklin Jones

*Professor*
*Dept. of Math and Computer Science*
*Vice President*
*Office of Institutional Effectiveness and*
*Sponsored Programs*
*Benedict College*
*joness@benedict.edu*

*Abstract*

*This paper discusses the real-time prediction of queue lengths from probe vehicles for the Bunch arrival headways at an isolated intersection for undersaturated conditions. The paper incorporates the bunching effect of the traffic into the evaluation of the accuracy of the predictions as a function of proportion of probe vehicles to entire vehicle population. Formulations are presented for predicting the expected queue length and its variance based on Negative Exponential and Bunched Exponential vehicle headways. Numerical results for both vehicle headway types are documented to show how prediction errors behave by the volume to capacity ratio and probe proportions. It is found that the Poisson arrivals generate conservative confidence intervals and demand higher probe proportions compared to Bunched Exponential headways at the same arrival rate and probe proportion.*

## 1. Introduction

Queue length is one of the fundamental performance measures of signalized intersections. Accurate prediction of such measures in real-time enable better control through efficiently allocating the available capacity (i.e., green time) such that a defined performance metric is optimized (e.g., minimize total delays or minimize the maximum queue length). To prediction these performance measures in real time, various surveillance technologies are being employed today (e.g., inductive loops, video) to measure traffic flow parameters (e.g., volume, density) which are subsequently utilized in models for delay prediction/prediction. Such systems are called real-time traffic-responsive or traffic-adaptive control systems [1]. One fundamental component of any traffic-adaptive signal control system is the prediction of delays and/or queue lengths at signalized intersection approaches. This paper focuses on queue length prediction at an isolated intersection in real-time based on data from probe vehicles (i.e., vehicles equipped with GPS and wireless communication technologies). Although probe vehicles are used as traffic data source for queue length prediction for Poisson arrivals in [2], this study differently contributes to the field via embedding the bunching effect of traffic into queue length prediction problem. The relationship among the accuracy of the predictions, the probe percentage (i.e., probe proportions, probe levels), arrival headways, and the arrival rates are explored.

Earlier studies on vehicle probes deal with understanding the relationships between the probe proportion and the reliability of the travel time or travel speed predictions [3-6]. Network coverage is also an important issue that is addressed in the literature [7-9]. Empirical analyses are performed in these studies based on data for numerous scenarios with different probe vehicle percentages. Typically, data from microscopic traffic simulation models are used since real-world data with a large number of probes to support such analyses are not available.

In this paper, analytical models are adopted from [2] to assess how queue length prediction is influenced by the percentage of probe vehicles in the traffic stream. These models require the marginal probability distribution of queue length to be known. The application of the proposed approach is illustrated through two examples for an isolated intersection with fixed signal timing. The arrivals are assumed to have the Bunched Exponential (Cowan's M3 Distribution) and the Negative Exponential headways whereas the vehicles are assumed to queue vertically for simplicity.

There are numerous vehicle headway studies in the literature. They can be categorized as single lane-isolated to multiple lane-series of intersections. The most famous of all, Cowan [10] investigated four different headway distributions. He concluded that the Bunched Exponential (M3) and its generalization M4 are more realistic for many traffic analyses. In another study, Buckley [11] compared several headway distributions (i.e., Negative Exponential, Displaced Exponential, Erlang, Gamma, Generalized Pearson Type III, and semi-random distributions) at different traffic volumes. The study obtains best fit semi-random distribution at high volumes and low volumes.

Also various other studies can be found in the literature regarding traffic headway models to express both signalized and unsignalized traffic flows [12-16].

The Bunched Exponential distribution is used by several researchers for its advantages over the Negative Exponential and the Shifted Exponential headways such as modeling the platoon effect in a traffic stream, generating smaller headways (i.e., high flow rates), and for its simplicity in generation random variates. For signalized intersections, Cowan [17] used the Bunched Exponential in an improved heuristic for actuated traffic signals. Similarly for actuated signal studies, Akcelik [18] used the Bunched Exponential distribution in green times and cycle time prediction models. In a very recent study, the distribution was deployed to evaluate green time extensions associated with two different vehicle detection schemes for actuated traffic signals [19].

Since fixed cycle traffic light allows a detailed analytical analysis, it has been studied by many researchers. The detailed review of the queue length prediction at fixed cycle traffic cycles can be found in [2, 20].

It is also interesting to note that the general idea of using probes to predict the system state or system performance is not unique to vehicular traffic on roadways. In computer communication networks, "probe packets" are sent from a source to one or more receiver nodes in the network in order to deduce the quality of service or performance (e.g., loss rate, delay) at the internal nodes or links (e.g., routers). In this technique, performance of the internal links/nodes is predicted by exploiting the correlation present in end-to-end (origin to destination) measurements obtained from probe packets (e.g., [21-22]).

This article is organized as follows. Section 2 introduces the problem statement and the notation. Section 3 describes the quick recall of analytical formulation for the queue length prediction problem. The application of the formulation is presented first for a fixed-cycle traffic light without considering the overflow queue in Section 4. Section 5 addresses the application of the formulation when overflow queue is also considered. Lastly, Section 6 summarizes the findings and results.

## 2. Problem Statement and Notation

Figure 1 illustrates a snapshot of a signalized intersection approach at the end of a red phase. Solid rectangles represent probe vehicles. The main objective is to predict the total queue length, $N$, if the locations of probe vehicles in the queue are known. Technically, the goal is to determine the conditional expected value of $N$ given the probe information (i.e., location of the last probe, $L_p$) and to evaluate the error when this conditional expected value is used as the predictor of the actual $N$, where $N$ is the total number of vehicles accumulated in the queue at the end of red interval.

It is assumed that the distance of probe vehicles from the stop bar can be measured by location tracking technologies. For example, today's differential GPS systems can provide one to three meter accuracy [23]. Once the position of a vehicle waiting on the link is known (i.e., distance from the stop bar), the number of vehicles ahead of the subject vehicle can be predicted by assuming an average length/spacing per vehicle. Even though all vehicles are assumed to be identical in this paper, it is possible to capture the impact of multiple vehicle classes through specifying arrival rates for each vehicle type. The information flow architecture and the exact nature of data processing needed to obtain the location of probes in the queue is not the scope this paper. Albeit important, such details as determining whether a probe vehicle is in the queue or not (i.e., waiting or moving) which can be determined based on its speed, acceleration and selected thresholds are not dealt with in this paper. Also for simplicity, the location of vehicles in queue will be measured in terms of the number of vehicles.

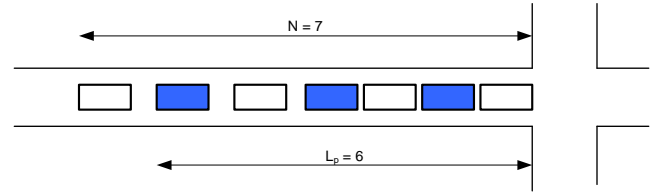

**Figure 1** Snapshot of an intersection right before the red interval terminates

## 3. Analytical Formulation

All of the formulations in this section are adopted from [2]. The section briefly summarizes them to clarify the approach to the problem. Assuming that the proportion of probe vehicles $p$ is known, the relationship between the total number of probe vehicles $N_p$ and the total number of all vehicles in the queue $N$ can be written as follows.

$$N_p = \sum_{i=1}^{N} y_i \quad \text{where } y_i \in \{0,1\} \text{ and } P(y_i = 1) = p \quad (1)$$

The equation above asserts that every vehicle has equal probability of being a probe vehicle. It is assumed that the random variables $y_1, y_2,\ldots$ are independent and independent of $N$. Both $N$ and $N_p$ are two discrete random variables. The formulation here is for a general probability mass function (pmf) for $N$, which is denoted by $P(N=n)$. As it is shown in [2], this probability only depends on the location of the last probe $l_p$ and does not depend on the total number of probes in the queue $n_p$. It should be noted that equation (2) is valid for any probability distribution of $N$. Even though the numerical examples presented in the next sections are based on Poisson and the count distribution of the Bunched

Exponential headways, the formulation can be applied for any given $P(N=n)$. The expected queue length computed as follows.

$$E(N=n|l_p) = \sum_{n=l_p}^{\infty} n P(N=n|l_p)$$

$$= \sum_{n=l_p}^{\infty} n \frac{[(1-p)]^n P(N=n)}{\sum_{k=l_p}^{\infty} [(1-p)]^k P(N=k)} \quad (2)$$

The equation above gives the mean of the random variable $N/L_p=l_p$ which can be used as a predictor of the queue length given the probe location information. For real-time applications, this equation can be used to prediction the queue length, provided that $P(N=n)$ is known.

In order to assess how accuracy of this predicted queue length changes by the proportion of probe vehicles $p$, some additional results are needed. First, the conditional variance of the queue length (given $l_p$) can be written as follows.

$$VAR(N=n|l_p) = \sum_{n=l_p}^{\infty} n^2 \frac{[(1-p)]^n P(N=n)}{\sum_{k=l_p}^{\infty} [(1-p)]^k P(N=k)} \quad (3)$$

$$- [E(N=n|l_p)]^2$$

Since the above variance depends on $l_p$, a more general measure that does not depend on $l_p$ is needed to determine the effects of $p$ on accuracy. The error in the predictions, denoted by $D$, can be treated as a random variable that is the difference between the actual queue length $N$ and its prediction:

$$D = N - E(N|L_p = l_p) \quad (4)$$

If the expectation of both sides of this equation is taken, it can be seen that the expected error is zero (Since $E[E(N|L_p=l_p)] = E(N)$). On the other hand, the variance of error is as follows.

$$Var(D) = Var[N - E(N|L_p=l_p)] = E[Var(N|L_p=l_p)] \quad (5)$$

The expected value on the right hand of this equation can be calculated as follows.

$$E(Var(N|L_p=l_p)) = \sum_{l_p=0}^{\infty} P(L_p=l_p) * VAR(N|L_p=l_p) \quad (6)$$

This expectation in essence represents the weighted variance over all possible values of $l_p$. To calculate this weighted variance, the marginal probability distribution of $L_p$ is needed. This marginal distribution can be readily written in terms of the conditional distributions obtained thus far. The marginal distribution of $L_p$ becomes,

$$P(L_p=l_p) = \sum_{n_p=1}^{l_p} \sum_{n=l_p}^{\infty} \binom{l_p-1}{n_p-1} p^{n_p}(1-p)^{n-n_p} P(N=n), \text{ for } l_p \geq 1 \quad (7)$$

$$P(L_p=0) = \sum_{n=0}^{\infty} (1-p)^n P(N=n), \text{ for } l_p = 0$$

The next two sections provide two examples to illustrate the application of these formulations. Both examples are repeated for Negative exponential and Bunched Exponential vehicle headway distributions.

## 4. Analysis of Arrivals on Red (without Overflow Queue)

In this first example, overflow queue (i.e., residual queue, initial queue at the end of green period) is omitted. The headway distributions are assumed follow the Negative Exponential distribution (i.e., the count distribution is Poisson arrivals) and the Bunched Exponential distribution. Poisson vehicle arrivals were first tested by [24]. He compared the actual arrivals with the arrivals from Poisson distribution and concluded that they are in good agreement at relatively low traffic volumes. However, in a later study showed that for less than two lanes it is inevitable for vehicles to form bunches [17]. Moreover, the Negative exponential distribution is unable to express small headways or high flow rates of traffic characteristics. Thus, Cowan's M3 or the Bunched Exponential distributions found to be a good alternative to overcome these shortcomings. The distribution used by several researches to express vehicle headways at uninterrupted and interrupted highways.

If the headways are assumed to follow the Negative Exponential distribution then the counting distribution or the number of arrivals per unit time is Poisson. Hence, $N$ follows Poisson distribution with parameter $\lambda$ per red duration. Then the probability function for $N$ can be written as,

$$P(N=n) = \frac{e^{-\lambda}(\lambda)^n}{n!}, \lambda > 0, n = 0,1,2... \quad (8)$$

For illustration purposes an example is constructed where $\lambda$ is set to 10 for red duration. From equation (2), the conditional expected queue length for the Negative Exponential headways can be obtained as follows.

$$E(N|L_p=l_p) = \sum_{n=l_p}^{\infty} n \frac{[(1-p)\lambda]^n}{n! \sum_{k=l_p}^{\infty} \frac{[(1-p)\lambda]^k}{k!}}, \text{ for } l_p \geq 1 \quad (9)$$

The Bunched Exponential headway distribution describes the traffic flow as consisting of two components; the first component is the bunched (restricted) vehicles that are tracking at an assumed constant headway $\Delta$, while the free vehicles are traveling at headways greater than $\Delta$.

$$P(T \leq t) = \begin{cases} 1 - \phi \exp[-\theta(t - \Delta)] & \text{for } t \geq \Delta \\ 0 & \text{for } t < \Delta \end{cases} \quad (10)$$

Where $\phi = \exp(-b\Delta\lambda)$ is the proportion of free vehicles to entire traffic and the parameter $\theta$ is given by $\theta = \phi\lambda/(1-\Delta\lambda)$. Parameters $\Delta = 1.5$ and $b = 0.6$ (i.e., for single-lane case) are adopted from [16].

$$P(R = r) = \phi(1-\phi)^{r-1} \quad \text{for} \quad r \geq 1 \quad (11)$$

Bunch size, r distribution is assumed to have the Geometric distribution as shown in equation (11). The Geometric distribution was tested in [10, 25-26] along with the Borel-Tanner and the Miller distributions. It was found to be a good fit for bunch sizes particularly for single-lane no overtaking case. Thus, the distribution of *N* for bunch arrivals, *P(N=n)* is found with simulation and plugged into equation to predict *N given location of the last probe*. Then *VAR(D)* is calculated. Figure 2 shows the behavior of the error changing with the probe proportion levels for both headway distributions.

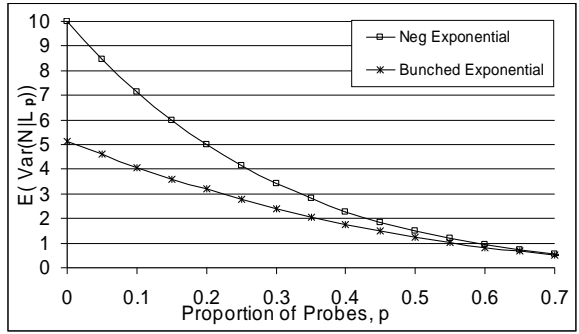

**Figure 2** Variance of Error versus p

Even though Figure 2 provides some guidance on trading off accuracy and *p*, confidence intervals are more informative for making such decisions. In order to construct confidence intervals for *D* at a given *p* value, one needs to know the pmf of *D* shown in equation (4). However, deriving this pmf seems to be prohibitive due to the complexity of the expressions involved (e.g., the probability distribution of $E(N/L_p=l_p)$). However, one can always use a probability inequality such as Vysochanskii-Petunin inequality to show some practical implications. The inequality holds for any random variable from any unimodal distribution.

$$P(|X - \alpha| > \varepsilon) \leq \begin{cases} \dfrac{4\xi^2}{9\varepsilon^2} & \text{for all } \varepsilon \geq \xi\sqrt{8/3} \\ \dfrac{4\xi^2}{9\varepsilon^2} - \dfrac{1}{3} & \text{for all } \varepsilon \leq \xi\sqrt{8/3} \end{cases} \quad (12)$$

Where $\xi^2 = E(X-\alpha)^2$ for an arbitrary $\alpha$, taking $\alpha = \mu$ and $\varepsilon = 3\sigma$ where $\sigma^2 = \xi^2$. The inequality yields $P(|X - \mu| > 3\sigma) \leq 4/81 < 0.05$. This is the so-called three-sigma rule widely used in quality control.

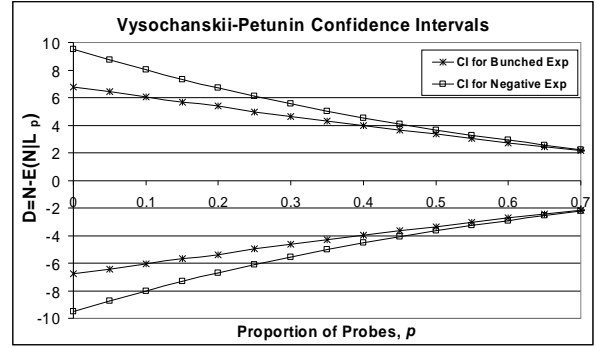

**Figure 3** Confidence Intervals (CIs) for Negative Exponential and Bunched Exponential Headways

The previous figures are constructed for a fixed arrival rate $\lambda$ which essentially determines the average queue length. In order to see how *VAR (D)* behaves with different arrival rates for the Bunched Exponential headways, Figure 4 is constructed which shows the error normalized by the arrival rate $\lambda$ in percentage as a function of *p* at various $\lambda$ levels. At a given *p* level, the relative or normalized error is smaller for larger $\lambda$ values which implies that the relative error gets smaller as average queue length increases.

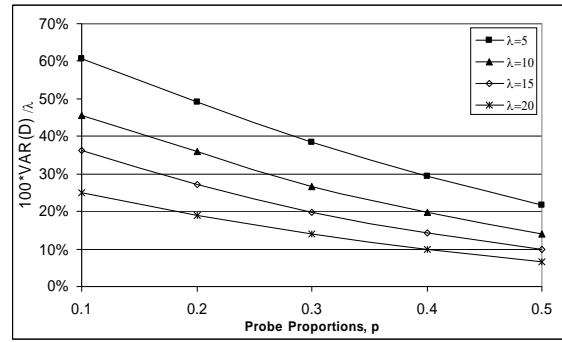

**Figure 4** Confidence Intervals (CIs) for Negative Exponential and Bunched Exponential Headways

Figure 5 compares the two headway distributions for different arrival rates (i.e., 400-1600 vph). Intuitively, one can expect bunching to diminish as the arrival rate reduces. Figure 5, agrees with this phenomenon.

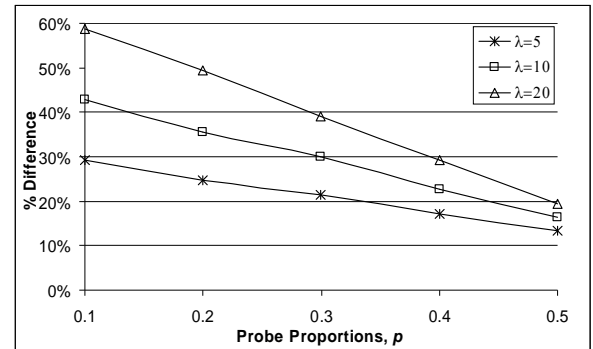

**Figure 5** Confidence Intervals (CIs) for Negative Exponential and Bunched Exponential Headways

## 5. Analysis with the Overflow Queue

In this section, the total queue length prediction that also includes the overflow queue due to randomness is presented. It is still assumed that average demand is less than the capacity – undersaturated conditions. The main input needed is the marginal probability of the total queue length, $P(N=n)$. For a traffic light with fixed timing, some researchers developed models that allow calculation of $P(N=n)$ numerically. These models involve finding the complex roots of the pgf of overflow queue in equilibrium [27]. The application of these models also requires numerical evaluation and approximations for pgf inversion [28]. The second alternative to find $P(N=n)$ is through simulation. In this paper, a simulation model is developed to determine $P(N=n)$.

In this example, the vehicles are assumed to depart from the stopbar at a constant rate during the green, which is assumed to be 2 seconds per vehicle. Therefore, the service rate $\mu$ is equal to the green duration divided by 2. The distribution of total queue length, $P(N=n)$, is determined by multiple simulation runs. The traffic light parameters are chosen as follows: cycle length, $C= 90\ s$; red period, $r= 45\ s$; and green period, $g= 45\ s$. The arrival rate, $\lambda$, is set to 20 vehicles per cycle. These selected parameters correspond to a $v/c$ ratio of 0.88. Simulation is run for 65,000 cycles. The first 200-cycles in each replication is considered as the warm-up period and not included in calculating the equilibrium distribution of $P(N=n)$.

Once the $P(N=n)$ is obtained, an analysis similar to the one presented in the previous section can be performed to determine the expected total queue length and variance.

**Table 1** 3σ deviations at various $p$ levels

| Probe Propor, $p$ | $10^{(-4)}$ | 0.1 | 0.2 | 0.3 | 0.4 | 0.5 |
|---|---|---|---|---|---|---|
| # of Vehs Bnch(3σ) | 6.9 | 6.2 | 5.4 | 4.7 | 4.0 | 3.4 |
| # of Vehs Neg Exp(3σ) | 12.8 | 9.8 | 7.8 | 6.2 | 4.9 | 3.9 |

In Table 1, absolute 3σ deviations are presented for both Bunched Exponential and the Negative Exponential distributions. It is useful to look at absolute deviations in terms of the number of vehicles. For example, for 10% probe vehicles, the absolute error is about ±6 vehicles for the Bunched Exponential and ±10 vehicles for the Negative Exponential. Thus, the assumption of Poisson arrivals demands higher $p$ level to be in ±6 at $\rho$=0.88. The difference between absolute errors reduces as the probe level increases. These deviations roughly correspond to a 95% confidence interval if $D$ follows any unimodal distribution.

Thus far, the results presented are for a constant arrival rate with a $v/c$ ratio or $\rho$ equal to 0.89. To understand the impact of $\rho$ on accuracy, the arrival rate is varied to generate the scenarios shown in Table 2 where the percent of probe vehicles $p$ is kept constant at 50%. For each one of these scenarios, the simulation program is run separately to obtain $P(N=n)$. Last two columns of Table 2 convey the 3σ percent deviations from the mean arrival rates. As it can be observed, absolute deviations are not increasing significantly as $v/c$ increases. On the other hand, percent deviations from the mean are decreasing substantially with increasing $\rho$. This can be attributed to the increase in the size and variance of the initial queue.

**Table 2** Results for various $\rho$ values

| p=0.5 | | Bnch Exp | Neg Exp | Bnch Exp | Neg Exp |
|---|---|---|---|---|---|
| $\rho=\lambda/\mu$ | $\lambda$ | $\sigma^2=E[VAR(N|L_p)]$ | $\sigma^2=E[VAR(N|L_p)]$ | 3σ Dev. % | 3σ Dev. % |
| 0.60 | 13.50 | 1.106 | 1.395 | 47% | 52% |
| 0.70 | 15.75 | 1.161 | 1.472 | 41% | 46% |
| 0.80 | 18.00 | 1.210 | 1.553 | 35% | 40% |
| 0.88 | 20.00 | 1.257 | 1.654 | 30% | 34% |
| 0.90 | 20.25 | 1.263 | 1.672 | 29% | 33% |
| 0.95 | 21.38 | 1.308 | 1.766 | 23% | 27% |
| 0.99 | 22.00 | 1.340 | 1.876 | 15% | 17% |

Figure 6 below shows that the headway distribution not only effects the decision on determining the appropriate probe level but also the decision of omitting or not omitting the overflow queue for a given arrival rate. For the Negative Exponential headways, omitting the overflow queue makes the power of the hypothesis tests less and requires more sample sizes (number of cycles) for the same confidence level since the $VAR(D)$ is substantially different at some points (e.g., 45-20% at p=0.0001-0.3). On the other hand for the Bunched Exponential headways, the difference is not more than 5% at any probe level.

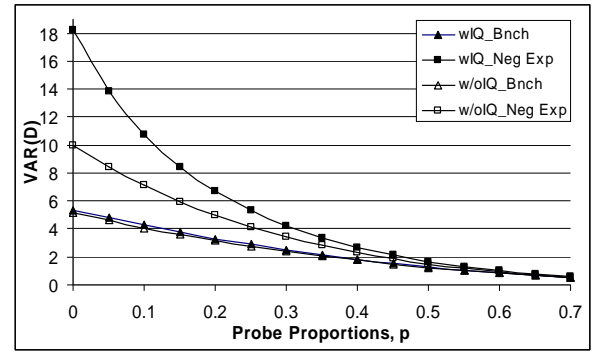

**Figure 6** VAR(D) for tow different headway distributions with and without considering overflow queue (IQ)

## 6. Conclusions

This paper presents a statistical method for real-time prediction of queue length at a signalized intersection approach from probe vehicle data under the Bunched Exponential vehicle headways and its comparison with Poisson arrivals. It is found that the formulations with Negative Exponential headways give conservative results compared to the Bunched Exponential distribution at higher flow rates and especially at low levels of probe penetrations.

Although the analytical formulations are straightforward, the decision tool of accuracy changing with probe proportion suggests costlier probe levels at the same level of precision. Analysis with overflow queue reveals that the Negative Exponential headways offers even more conservative results compared the Bunched Exponential. Obviously, this conclusion is valid for the conditions and assumptions considered in the paper (e.g., steady-state conditions, known arrival rate, and undersaturated conditions ($\rho < 1.0$)).

The major contribution of this paper is that the analysis here can be used for urban traffic settings. The results show that after about 40-50% probe penetration, prediction of $N$ can be carried out with the formulations of Poisson arrivals. In addition for arrival rate less than 10 vehicles per 90 seconds, $VAR(D)$ values are getting very close for both headway distributions.

## References

[1] Gartner N. H, F. J. Pooran, C. M. Andrews. Optimized Policies for Adaptive Control Systems: Implementation and Field Testing. Transportation Research Record 2002; No. 1811; 148-156.

[2] Comert, G., Cetin, M., Queue Length Prediction from Probe Vehicle Location and the Impacts of Sample Size. Eur. J. of Operational Research, 2008, doi:10.1016/ejor.2008.06.024.

[3] Chen M, Chien S.I.J. Determining the number of probe vehicles for freeway travel time prediction by microscopic simulation. Transportation Research Record 2000; No. 1719; 61-68.

[4] Cheu R. L, C. Xie, D. Lee. Probe Vehicle Population and Sample Size for Arterial Speed Prediction. Computer-Aided Civil and Infrastructure Engineering 2002; 17(1); 53-60.

[5] Ferman, M. A., Blumenfeld, D. E., Dai, X. An Analytical Evaluation of a Real-Time Traffic Information System Using Probe Vehicles. J. of Intelligent Tran. Systems: Technology, Planning, and Operations, 2005, 9 (1), pp. 23-24.

[6] Lin I, H. Rong, A. L. Kornhauser. Estimating Nationwide Link Speed Distribution Using Probe Position Data. Journal of Intelligent Transportation Systems Jan 2008; Vol. 12; 1; 29-37.

[7] Srinivasan K.K, P.P. Jovanis. Determination of Number of Probe Vehicles Required for Reliable Travel Time Measurement in Urban Network. Transportation Research Record 1996; No.1537; 5-22.

[8] Turner S, D. Holdener. Probe Vehicle Sample Sizes for Real-Time Information: The Houston Experience. Proceedings of the Vehicle Navigation & Information Systems Conference, Seattle, WA, United States; 1995; 3-10.

[9] Boyce D E, J. Hicks, A. Sen. In-Vehicle Navigation Requirements for Monitoring Link Travel Times in a Dynamic Route Guidance System. Operations Review 1991; 8 (1); 17-24.

[10] Cowan, R. Useful headway models. Transportation Res. 1975, 9, 371-375.

[11] Buckley, D. J. Road traffic headway distributions, in Proceedings, 1962, Vol. 1, part 1, Australian Road Research Board, Victoria, pp. 153–187.

[12] Gerlough, D.L. and Huber, M.J. Traffic Flow Theory. 1975, Special Report 165, Transportation Research Board, National Research Council, Washington, D.C.

[13] Lewis, R. M. A Proposed Headway Distribution for Traffic Simulation Studies. Traffic Engineering, February 1963, Vol. 33, 16-19, 48.

[14] Sullivan, D. P. and Troutbeck, R. J. The use of Cowan's M3 headway distribution for modelling urban traffic flow. Traffic Engineering and Control, 1994, 35(7/8), 445–450.

[15] Luttinen, R. T. Statistical Analysis of Vehicle Time Headways. Teknillinenkorkeakoulu, Liikennetekniikka, Julkaisu 87. Otaniemi, 1996. ISBN 951-22-3063-1.

[16] Akcelik, R. and Chung, E. Calibration of the bunched exponential distribution of arrival headways. Road and Transport Research 1994, 3 (1), pp 42-59.

[17] Cowan, R. An improved model for signalised intersections with vehicle-actuated control. J. Appl. Prob. 1978, 15, 384-396.

[18] Akcelik, R. Prediction of green times and cycle time for vehicle-actuated signals. Transportation Research Record 1457, 1994, pp 63-72.

[19] Tian, Z. and Urbanik, T. Green Extension and Traffic Detection Schemes at Signalized Intersections, Transportation Research Record 1978, 2006, pp. 16-24.

[20] Rouphail N., A. Tarko, J. Li. Traffic Flows at Signalized Intersections. Traffic Flow Theory Monograph; 2001. Chapter 9.

[21] Bowei X, G. Michalidis, V. N. Nair. Estimating Network Loss Rates Using Active Tomography. Journal of the American Statistical Association 2006; Vol. 101; No. 476; 1430-1448.

[22] Duffield N.G., Network Tomography of Binary Network Performance Characteristics, IEEE Transactions on Information Theory, 2006, vol. 52, pp. 5373-5388.

[23] USDOT, "Nationwide Differential Global Positioning System Program Fact Sheet," Turner-Fairbank Highway Research Center, http://www.tfhrc.gov/its/ndgps/02072.htm, visited on February 26, 2008.

[24] Adams, W. F. Road Traffic Considered as a Random Series. Institution of Civil Eng. J.,1936, 4.

[25] Chrissikopoulos, V., D., J. and McDowell, M. Aspects of headway distributions and platooning on major roads. Traffic Engineering and Control, 1982, 23(5), 268–271.

[26] Taylor, M. A. P., Miller, A. J. and Ogden, K. W. A Comparison of Some Bunching Models for Rural Traffic Flow. Transportation Res. 1974, 8, 1-9.

[27] Abate J, Whitt W. Numerical Inversion of Probability Generating Functions. Operations Research Letters 1992; 12; 245-251.

[28] Van Leeuwaarden V.J.S.H. Delay Analysis for the Fixed-Cycle Traffic-Light Queue, Transportation Science 2006; 40 (2); 189-199.